\documentclass[journal = jpcl,manuscript=letter,layout=letter]{achemso}
\usepackage{bm,graphicx,amsmath,amssymb,color,placeins,tikz,pgflibraryarrows,braket,colortbl,xcolor,mathtools}
\usepackage[caption=false]{subfig}
\usetikzlibrary{decorations.pathreplacing}

\def\<{\langle}
\def\>{\rangle}

\title{Local Structural Sisorder in Crystalline Materials}

\author{Marios Zacharias} 
 \email{zachariasmarios@gmail.com}
\affiliation{Computation-based Science and Technology Research Center, The Cyprus Institute, Aglantzia 2121, Nicosia, Cyprus}

\author{Jacky Even}
\affiliation{Univ Rennes, INSA Rennes, CNRS, Institut FOTON - UMR 6082, F-35000 Rennes, France}

\date{\today}

\begin{document}
\maketitle
\begin{abstract}
ABSTRACT:  Local positional disorder in soft, anharmonic materials has emerged as a central factor in shaping their electronic, 
vibrational, optical, and transport properties. Viewed mainly as a source of performance degradation, recent 
theoretical insights reveal that local disorder profoundly influences the electronic structure and phonon dynamics, 
 without inducing deep electronic traps or non-radiative recombination pathways.
In this Perspective, we highlight advances in modeling 
local disorder using polymorphous and anharmonic frameworks, showing how these methods explain experimental observations and 
predict new trends. We emphasize the role of disorder in the breakdown of the phonon quasiparticle picture and in 
modulating electron-phonon and phonon-phonon interactions, particularly in soft, anharmonic phases of matter, 
with significant effects on electrical and thermal transport. We outline opportunities 
for integrating these insights into predictive modeling for energy materials and propose combining 
advanced first-principles methods with machine learning.
\end{abstract}

\maketitle
%

\newpage

Local disorder is an intrinsic feature of soft, anharmonic materials, first realized in the late 60's by Comes et al. for 
oxide perovskites~\cite{Comes1968}. 
While standard diffraction analyses of crystalline materials suggest high-symmetry monomorphous structures,
local fluctuations in atomic positions give rise to a rich configurational landscape, as suggested by more 
advanced experimental approaches~\cite{Mashiyama1998,Worhatch2008,Beecher2016,Laurita2017,LaniganAtkins2021,
Dirin2023,Sabisch2025,Dubajic2025,Bhui2025,He2024,Chengjie2025}. 
These variations, linked to configurational entropy, are termed positional 
polymorphism~\cite{Zhao2020} and impact key physical properties in ways that are not captured by conventional, monomorphous models. 
Some communities refer to this effect as static disorder or split-site occupancy, terms that are related
but not strictly identical. Recent advances in first-principles theory have enabled the explicit treatment 
of such disorder~\cite{Zhao2020,Zacharias2023npj}, 
offering a new lens to interpret and predict material behavior in energy applications.

\begin{figure}[t!]
\includegraphics[width=0.75\textwidth]{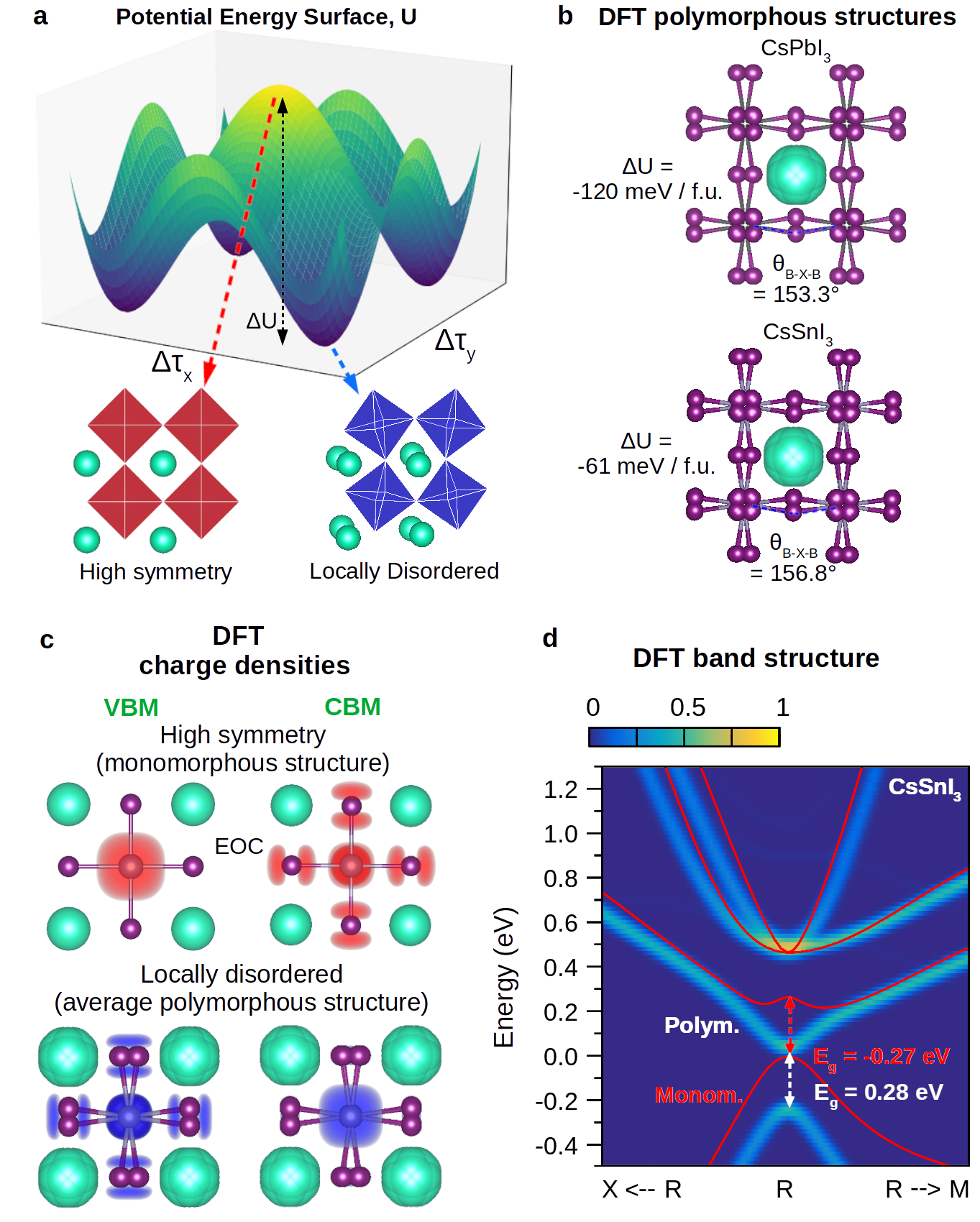}
\caption{\label{fig1}{\bf Impact of local disorder on the electronic structure.} (a) Schematic representation of the PES illustrating the high symmetry and locally disordered structures, and the corresponding energy lowering $\Delta U$. (b) Locally disordered (polymorphous) structures of cubic CsPbI$_3$ and CsSnI$_3$ obtained by the ASDM using a $4\times4\times4$ supercell and DFT geometry optimization. 
(c) Average DFT charge densities of high symmetry and locally disordered cubic CsSnI$_3$. 
(d) DFT electron spectral function of polymorphous
CsSnI$_3$ calculated by band structure unfolding~\cite{Zacharias2020}.
The band structure of the monomorphous model is shown as red. 
The band gap opening due to local disorder is $\Delta E_{\rm g} = 0.56$~eV.
 Data are from Refs.~\cite{Zacharias2023npj,Zacharias2025b}} 
\end{figure}

Local disorder is a typical characteristic of room temperature phases of materials, and particular of anharmonic 
materials featuring a multiwell potential energy surface (PES). The high symmetry configuration corresponds to a local maximum on 
the PES, and locally disordered configurations correspond to energetically favorable minima, as shown in Figure~\ref{fig1}a. 
At $T = 0$~K, both cases correspond to static equilibrium states. 
At higher temperatures, ultraslow relaxation occurs between locally disordered configurations. The polymorphous 
framework does not explicitly model these dynamics, but instead represents structural disorder through optimized 
supercells or a limited sampling of them.
This raises a fundamental question: which structural configurations 
should be considered as physically realistic for {\it ab initio} electronic structure calculations, particularly 
those based on density functional theory (DFT)? 
This choice also has profound implications for the coupling of carriers with lattice dynamics, i.e. 
electron-phonon coupling~\cite{Giustino2017}, as well as for phonon-phonon scattering which determine key optical and 
transport properties that govern the performance of energy devices.

In this Perspective, we present compelling evidence for the necessity of using locally disordered structures 
to achieve accurate predictions of electronic structure, vibrational dynamics, and electron-phonon interactions. Such an 
approach is crucial for capturing temperature-dependent properties, including vibrational spectra, effective masses, 
band gaps, and carrier mobilities, where theory results demonstrate excellent agreement with experimental data. 
We provide an analysis of ongoing theoretical 
and experimental efforts addressing the role of local disorder in materials, as well as outline limitations of the framework and 
emerging future directions. We discuss how local disorder can lead to 
improved predictions of key phenomena such carrier and thermal transport as well as polaron formation.
We believe this Perspective will contribute to a deeper understanding of the intricate interplay between structural disorder, 
electronic structure, anharmonicity, and electron-phonon coupling, while also accelerating the development of 
more accurate theoretical frameworks for predicting energy-relevant properties of functional materials.
Among the numerous materials known to feature local 
disorder~\cite{Comes1968,Trimarchi2018,Varignon2019,Zhao2021,Wang2025,Xiong2025,Bhui2025,Jakob2025_disorder}, 
emphasis is given to halide perovskites~\cite{Mashiyama1998,Zhao2020,Zacharias2025a}, a rapidly advancing area in energy research. 

\section{Electronic structure}

Local disorder in materials refers to spatially correlated deviations of atoms from their high symmetry positions, which on average
preserve the crystallographic space group symmetry of the structure~\cite{Zhao2020}. 
To simulate positional polymorphism, a natural approach is to employ {\it ab initio} 
molecular dynamics (MD) with extended equilibration and production times to sample the relevant configurational space. Another
approach is to start from a high symmetry supercell structure and 
introduce atomic displacements via random nudges~\cite{Zhao2020} or special displacements along 
phonons~\cite{Zacharias2020,Zacharias2023npj}. The structure is then relaxed with the lattice constants fixed. 
This method enables the system to explore symmetry-breaking domains and identify energetically favorable minima in the PES, 
as illustrated in Figure~\ref{fig1}a. The energy lowering with respect to the monomorphous structure is indicated as  
$\Delta U$. 

Figure~\ref{fig1}b shows the average DFT polymorphous structures explored within a $4\times4\times4$ supercell, 
using the example of Pm$\bar{3}$m cubic halide perovskites of type ABX$_3$: CsPbI$_3$ and CsSnI$_3$. The 
atomic positions in the supercell are folded back into the unit cell and symmetrized to reflect the underlying 
isotropy of the system. In each case, new structural degrees of freedom are revealed, where four equivalent 
iodine atoms are uniformly distributed around the original 
high symmetry site and lie on a plane perpendicular to the direction of the B-B axis. This is consistent with 
X-ray scattering diffraction measurements for cubic MAPbX$_3$ (X$=$Cl, Br, I) by Mashiyama et al.~\cite{Mashiyama1998}  
and cubic FASnI$_3$ nanocrystals by Dirin et al.~\cite{Dirin2023}, where MA and FA stand for methylammonium and formamidinium. 
These findings~\cite{Mashiyama1998,Dirin2023} along with the DFT-calculated polymorphous structures 
do not support the presence of local disorder due to metal off-centering, in contrast to Refs.~\cite{Laurita2017} 
and~\cite{Balvanz2024} which suggest a prominent Sn off-centering displacement in FASnX$_3$ (X$=$I, Br) due 
to the enhanced Sn lone pair activity. We note that further calculations using higher-level hybrid functionals, 
such as HSE, may prove useful in shedding additional light on this contradiction.
For this purpose, Ge-based compounds may serve as a better starting point to explore numerically the impact of polar 
configurations~\cite{Bhui2025}, as the 4s$^2$ lone pair of Ge$^{2+}$ 
exhibits stronger stereochemical activity than the 5s$^2$ or 6s$^2$ lone pairs of Sn$^{2+}$ and Pb$^{2+}$.

Focusing on non-polar fluctuations, the degree of local disorder is quantified by calculating the deviation of the B-X-B angle 
($\theta_{\rm B-X-B}$) from the ideal value of 180$^\circ$ in the cubic structure~\cite{Zacharias2025b}. 
In the examples shown in Figure~\ref{fig1}b, local disorder is greater in CsPbI$_3$ ($\Delta \theta_{\rm B-X-B}$ = 26.7$^\circ$) 
than in CsSnI$_3$ ($\Delta \theta_{\rm B-X-B}$ = 23.2$^\circ$), which is also reflected in the corresponding $\Delta U$ values for 
each structure. 
Due to stronger orbital overlap and enhanced bond covalency, CsPbI$_3$ has a larger lattice constant 
(6.25~\AA~vs. 6.14~\AA~in CsSnI$_3$), which in turn provides greater structural flexibility and a higher tendency toward 
local symmetry breaking. As discussed in Ref.~\cite{Zacharias2025a}, another key parameter 
determining the degree of local disorder in cubic halide perovskites is the tolerance factor. This is strongly influenced 
by the size of the A-site cation, with higher values 
indicating less local distortions. For example, FA-based compounds have been shown
to exhibit the lowest degree of positional polymorphism due to larger tolerance factors than MA-based and Cs-based 
compounds~\cite{Zacharias2025b}. Aziridinium (AZ)-based halide perovskites, a recent addition to the family of A-site 
cations~\cite{Zheng2018,Bodnarchuk2024}, are expected to exhibit a degree of local disorder in the range 
($\Delta \theta_{\rm B-X-B}$ = 13--18$^\circ$) higher than FA-based compounds ($\Delta \theta_{\rm B-X-B}$ = 9--12$^\circ$), 
yet slightly lower than that typically observed in MA-based systems ($\Delta \theta_{\rm B-X-B}$ = 14--19$^\circ$)~\cite{Zacharias2025a}. 
This is because the estimated ionic radius of AZ is approximately 2.30~\AA, which is intermediate between FA (2.53~\AA) 
and MA (2.17~\AA); the degree of local disorder in AZ-based compounds remains to be confirmed by first-principles calculations.

Figure~\ref{fig1}c shows that using a monomorphous model in DFT calculations for cubic CsSnI$_3$ and CsPbI$_3$, including 
the effect of spin-orbit coupling (SOC), leads to an unphysical result. 
The charge density at the valence band maximum (VBM) and conduction band 
minimum (CBM) exhibits a spurious exchange in orbital character, with the VBM formed primarily by B-site metal p 
orbitals and the CBM showing contributions from X-site halogen p states and B-site s states. This results in 
a spurious negative gap of -0.27~eV and negative electron effective masses at the R-point 
for CsSnI$_3$ as shown in Figure~\ref{fig1}d.
In contrast to common beliefs, this is not only a deficiency of DFT to handle electron correlations but a consequence of 
using the incorrect structural model without local disorder. As shown in Figure~\ref{fig1}c for the average polymorphous structure, 
the expected orbital character in the charge density at the band extrema is recovered, leading to an overall 
band gap opening of 0.56~eV (Figure~\ref{fig1}d). This value correlates with the B-X-B bond angle changes
associated with local disorder and tolerance factor, consistent with trends calculated for other halide perovskites~\cite{Zacharias2025a}. 
Accounting for local symmetry breaking is essential to recover reliable band gap behavior, as also shown for
oxide perovskites~\cite{Varignon2019}, low-dimensional 
quantum materials~\cite{Xiong2025}, transition metal monoxides~\cite{Trimarchi2018}, transition metal chalcogenides~\cite{Goesten2022}, 
and Cu-based superionic conductors~\cite{Wang2025}. 
We stress that while DFT combined with locally disordered structures captures key qualitative features, accurate band gaps 
and effective masses 
require advanced methods like hybrid functionals or $GW$ to account more accurately for electron correlation~\cite{Sutton2018,Zacharias2025b}.

\section{Structural and Vibrational properties}

\begin{figure}[t!]
\includegraphics[width=0.95\textwidth]{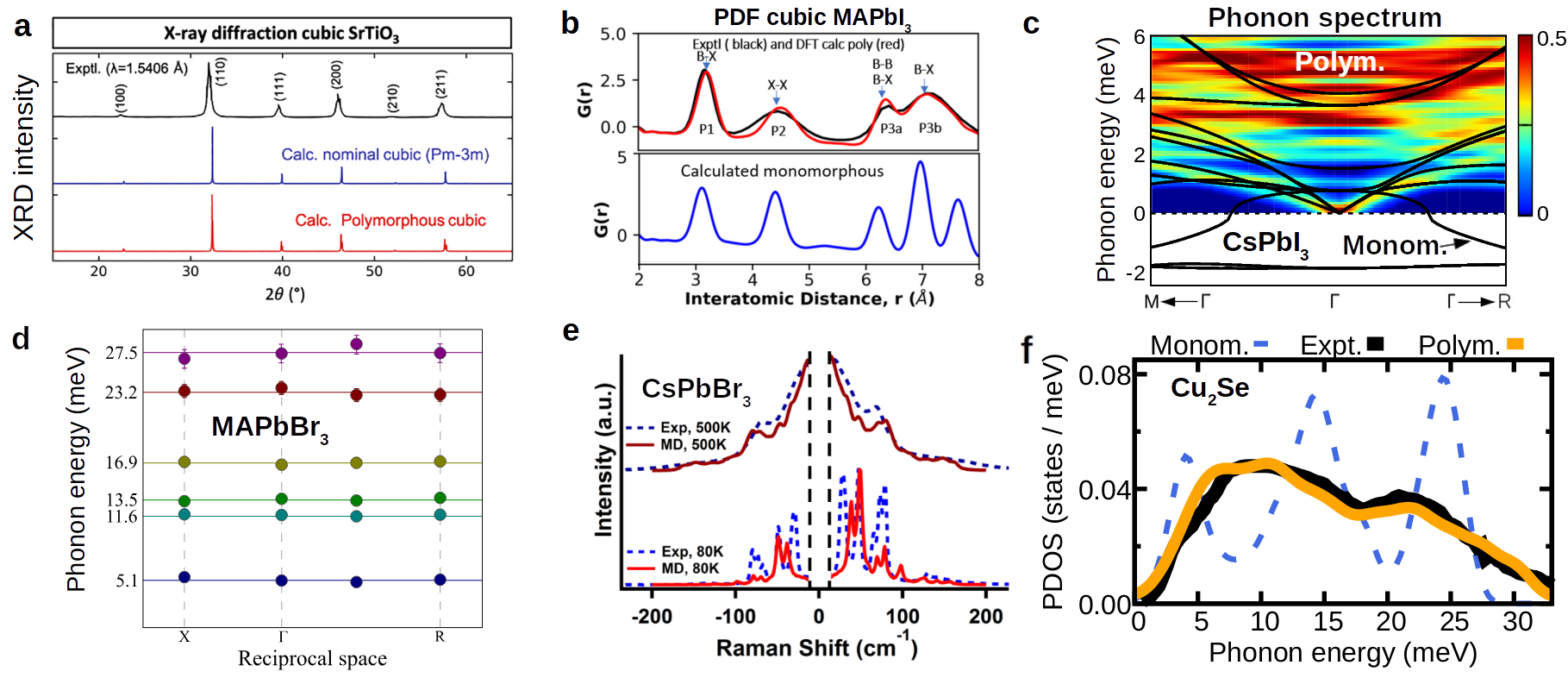}
\caption{\label{fig2} {\bf Impact of local disorder on structural and vibrational properties.} (a) Calculated XRD patterns 
of monomorphous and polymorphous cubic SrTiO$_3$ compared to experiment. 
Reproduced from Ref.~\cite{Zhao2021}, copyright (2021), with permission from Elsevier.
(b) Calculated PDF of monomorphous and polymorphous cubic MAPbI$_3$ compared with experiment. 
Reproduced with permission from Ref.~\cite{Zhao2020}, copyright (2020) by the American Physical Society.
(c) Phonon spectral function of polymorphous CsPbI$_3$ obtained by phonon unfolding. Data extracted from Ref.~\cite{Zacharias2023npj}.
(d) Phonon energies in reciprocal space measured for MAPbBr$_3$ with inelastic neutron scattering. Figure adapted 
from Ref.~\cite{Ferreira2020} under CC BY 4.0.
(e) Calculated Raman spectra of CsPbBr$_3$ by MD compared to experimental measurements. 
Reproduced with permission from Ref.~\cite{Yaffe2017}, copyright (2017) by the American Physical Society.
(f) Calculated phonon DOS of monomorphous and polymorphous cubic Cu$_2$Se compared with experiment. Data from Ref.~\cite{Wang2025}.
}

\end{figure}

Figure~\ref{fig2} illustrates the critical role of polymorphism in capturing the structural and vibrational 
properties of anharmonic materials. In Figure~\ref{fig2}a, Zhao et al.~\cite{Zhao2021} demonstrate using cubic 
SrTiO$_3$ that both monomorphous and polymorphous structures yield nearly identical X-ray diffraction (XRD) patterns, 
despite their underlying structural differences. This result is expected, as XRD primarily probes the average long-range order 
of atoms and is therefore not the most appropriate technique to reveal local deviations that preserve the overall 
crystallographic symmetry. A similar conclusion is reflected in Figure~\ref{fig1}d, where the band edges of the polymorphous 
structure remain doubly degenerate and exhibit no Rashba splitting, consistent with the preservation of inversion symmetry.
An experimental approach, sensitive to short-range order and local distortions, is measuring the pair distribution function (PDF)
typically via wide-angle X-ray or neutron scattering~\cite{Dirin2023}. Figure~\ref{fig2}b shows that the calculated PDF 
of polymorphous cubic MAPbI$_3$~\cite{Zhao2020} obtained from static DFT is in excellent agreement with experimental data 
at 290~K, reported in Ref.~\cite{Beecher2016}. In contrast, the monomorphous Pm$\bar{3}$m structure fails to reproduce 
experiment, both in terms of peak positions and broadening features.

Positional polymorphism also strongly affects the phonon spectra, leading to strongly overdamped
vibrational dynamics, as shown for CsPbI$_3$ by Zacharias et al.~\cite{Zacharias2023npj} (color map in Figure~\ref{fig2}c). 
Employing a monomorphous structure together with the harmonic approximation to compute phonons 
results in pronounced dynamical instabilities in the phonon dispersion, as shown by the black lines in Figure~\ref{fig2}c. This 
occurs because the monomorphous configuration corresponds to a local maximum on the PES, leading to negative curvatures 
along certain vibrational modes, manifested as imaginary phonon frequencies.
Instead, using a polymorphous network yields dynamically stable phonons, with acoustic branches emerging near 
the $\Gamma$-point from a background of nearly dispersionless optical vibrations, consistent with inelastic neutron 
scattering measurements in halide perovskites~\cite{Ferreira2020}, shown in Figure~\ref{fig2}d for MAPbBr$_3$. 
The strongly overdamped vibrations lead to the breakdown of the conventional phonon 
quasiparticle picture that is typically assumed in the monomorphous model. 
This breakdown becomes even more pronounced in hybrid halide perovskites due to enhanced mode mixing between polar 
optical phonons and the internal vibrations of the MA or FA molecules~\cite{Zacharias2025b}.
We remark that in many systems, unstable modes become anharmonically stabilised at higher temperatures, still 
retaining a well-defined quasiparticle picture as shown by MD-based TDEP calculations~\cite{Klarbring2020}.

Raman scattering measurements on halide perovskites reported by Yaffe et al.~\cite{Yaffe2017} reveal 
strongly correlated vibrational dynamics, as evidenced by a pronounced central peak spanning a broad 
frequency range, as shown in Figure~\ref{fig2}e. This feature is absent in monomorphous harmonic systems, which 
typically exhibit sharp and well-defined Raman peaks. Interestingly, MD can reproduce the 
central peak of the high temperature phase of CsPbBr$_3$, demonstrating that MD can capture both thermal vibrations and local disorder 
effects. Furthermore, MD has been proven powerful to capture overdamped phonon spectra 
in halide perovskites by analysis of velocity autocorrelation functions~\cite{Lahnsteiner_2025}.
Another observable that can be used to identify strongly overdamped, anharmonic vibrations is 
the phonon density of states (DOS). Figure~\ref{fig2}e shows the phonon DOS of Cu$_2$Se, a
polymorphous material with high degree of local disorder, as calculated by Wang et al.~\cite{Wang2025} using the 
anharmonic special displacement method (ASDM)~\cite{Zacharias2020,Zacharias2023}. The presence of positional polymorphism 
leads to significant spectral broadening, in good agreement with experiment~\cite{Liu2016}, and contrasts with the
well-defined peaks calculated for the monomorphous structure. 
This behavior is consistent with prior observations of strongly overdamped vibrational spectra in 
superionic materials~\cite{Gupta2022}.
Oxide perovskites represent another class of materials where overdamped phonon dynamics have been 
measured and supported through MD simulations~\cite{He2025}.

The breakdown in the phonon quasiparticle picture, apart from resulting in broadened phonon spectra and reduced lifetimes, it also 
leads to enhanced phonon-phonon scattering, suppressing lattice thermal conductivity due to increased energy dissipation 
and reduced phonon mean free paths~\cite{Niedziela2020,Gupta2022,Wang2025,Bhui2025}. For thermoelectric materials, this suppression 
of thermal conductivity 
is beneficial, as it enhances the figure of merit, provided that the electronic transport remains relatively unaffected. Therefore, 
understanding and controlling disorder-induced phonon scattering is not only crucial for accurately modeling thermal transport but also 
offers a strategy to engineer high-performance thermoelectrics.

\section{Thermal fluctuations in an anharmonic lattice}
\begin{figure}[t!]
\includegraphics[width=0.95\textwidth]{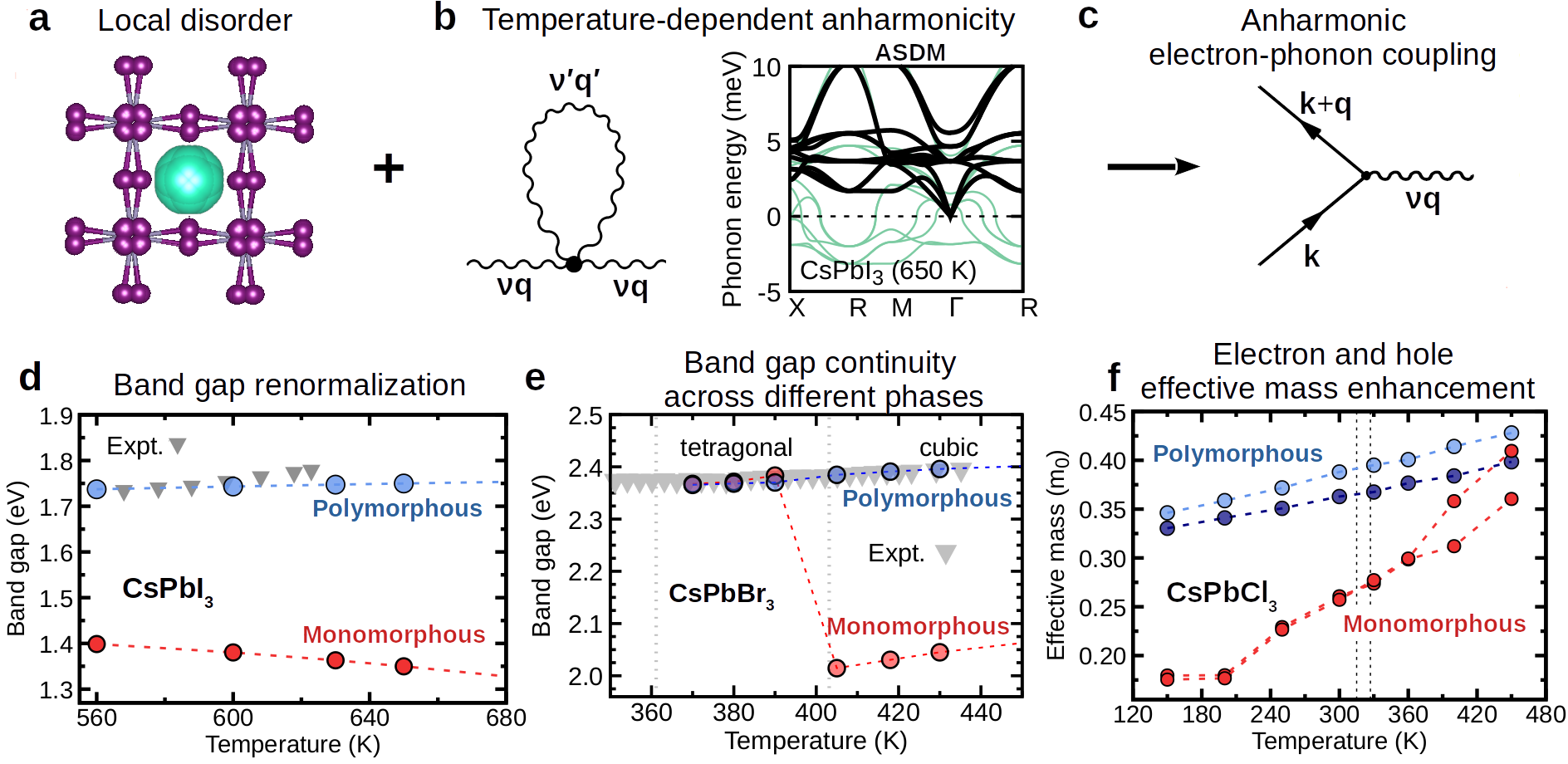}
\caption{\label{fig3} {\bf Impact of local disorder on electron-phonon coupling.} (a,b,c) Locally disordered structure and 
diagrammatic representation of loop diagram contributing to temperature-dependent anharmonicity and the basic electron-phonon 
coupling process. Lines represent electrons with wavector ${\bf k}$ and ${\bf k + q}$ and the zigzag lines
the anharmonic phonon. (d,e) Calculated band gap renormalization of monomorphous and polymorphous cubic CsPbI$_3$ 
and tetragonal/cubic CsPbBr$_3$ by ASDM compared with experiment. Data taken from Refs.~\cite{Zacharias2025b,Zacharias2023npj}. 
(f)  Calculated electron and hole effective mass of monomorphous and polymorphous cubic CsPbCl$_3$. 
Data extracted from Ref.~\cite{Zacharias2025b}. Vertical dashed lines indicate temperatures where phase 
transitions take place.}
\end{figure}

In Figures~\ref{fig3}{\bf a} and {\bf b}, we illustrate the two sources of anharmonicity that influence the vibrational 
dynamics: one is positional polymorphism, as discussed for Figure~\ref{fig2}, and the other arises from temperature-dependent 
phonon self-energy corrections. In the context of lattice dynamics, the loop diagram originating from quartic anharmonicity 
(Figure~\ref{fig3}{\bf b}) represents the first correction to phonon frequencies beyond the harmonic approximation~\cite{Tadano_2015}. 
It corresponds to a process where a phonon interacts with the thermal fluctuations of the lattice, 
mediated by a virtual phonon through a quartic vertex. Although quartic terms are associated with 
the 4th-order potential, the loop diagram does not represent a real scattering process among four phonons. Instead, it 
captures a self-interaction mechanism, where thermal fluctuations in the lattice feed back into the mode's own dynamics. 
This correction affects only the real part of the phonon self-energy and introduces no temperature-dependent broadening 
or lifetime effects, usually obtained by the so called bubble diagram~\cite{Tadano_2015}.
Within the self-consistent phonon (SCP) theory, temperature-dependent phonons are computed by iteratively solving an 
effective Hessian derived from interatomic forces. To obtain the final renormalized phonon frequencies and eigenvectors, 
SCP theory effectively resums anharmonic loop contributions to infinite order, offering a physically grounded and 
non-perturbative description of strong anharmonicity.

There are several state-of-the-art methods and codes that implement the SCP theory and extensions beyond it, including the
Temperature Dependent Effective Potential ({\tt TDEP})~\cite{Hellman_2011}, {\tt ALAMODE}~\cite{Tadano_2015}, stochastic 
self-consistent harmonic approximation ({\tt SSCHA})~\cite{Monacelli2021}, and the {\tt ASDM}~\cite{Zacharias2023}.
Within the same level of approximation, such as consistent calculation settings and order of anharmonicity, 
these methods are expected to yield identical results.
We emphasize that these approaches to temperature-dependent anharmonicity are conceptually 
different than the polymorphous framework. They primarily incorporate phonon self-energy corrections either using 
thermally displaced configurations generated by MD, Monte Carlo, or special displacements
sampling, yielding renormalized frequencies and linewidths, while preserving the average crystal symmetry and therefore do not 
explicitly explore local symmetry breaking. A key strength of these methods is the systematic use of crystal symmetries, which 
accelerates convergence and enables extensions beyond the loop diagram, for example, to capture corrections involving third- 
and higher-order anharmonicity~\cite{Bianco2017,Klarbring2020}. In contrast, the polymorphous framework focuses on explicitly 
capturing local symmetry breaking and positional polymorphism. 
Phonon self energy corrections can be incorporated on top of the polymorphous framework starting from optimized 
supercells~\cite{Zacharias2023npj}.

\section{Electron-phonon coupling}

The anharmonic phonons obtained by accounting for local disorder, phonon self-energy corrections, or both can be coupled with 
the electronic structure of the locally disordered system, yielding the so called anharmonic electron-phonon 
coupling (Figure~\ref{fig3}{\bf c}). Electron-phonon coupling, in turn, affects both phonon and electron energies and 
lifetimes through self-energy corrections, as discussed extensively in Ref.~\cite{Giustino2017}.
Figure~\ref{fig3}{\bf d} shows the effect of electron-phonon coupling on the band gap renormalization of 
cubic CsPbI$_3$, as reported in Ref.~\cite{Zacharias2025b}. DFT calculations are performed within the ASDM using 
a 4$\times$4$\times$4 supercell. Results are averaged over 10 polymorphous configurations, weighted by their Boltzmann factors
derived from free energies, thereby incorporating configurational entropy effects into the electron-phonon calculations. 
Employing polymorphous structures yields a positive band gap temperature coefficient, i.e., a band gap opening with 
increasing temperature.  In contrast, calculations based on a monomorphous structure, not only underestimate the band gap, 
but also predict a band gap closing, due to an incomplete treatment of the Fan-Migdal self-energy contribution~\cite{Zacharias2025a}.
In both cases, band gap renormalization is dominated by low energy optical modes ($<$10 meV) with bending or rocking character, 
linked to B-X-B bond angle variations and octahedral tilting~\cite{Zacharias2025a}. 

Polymorphous networks are also essential for describing the continuity of the band gap across structural 
phase transitions, as shown in Figure~\ref{fig3}{\bf e} for CsPbBr$_3$. Using a monomorphous model, a spurious 
band gap drop is observed when transitioning from the tetragonal to the cubic phase, caused by the enforced alignment of 
octahedra in the high symmetry Pm$\bar{3}$m structure. In contrast, employing a polymorphous network yields a smooth increase 
in the band gap with temperature, as local atomic disorder is retained in both phases. Furthermore, the energy barrier associated 
with the phase transition is significantly lower when moving between polymorphous tetragonal and polymorphous cubic phases, compared 
to their monomorphous counterparts~\cite{Zacharias2023npj}. 
Figure~\ref{fig3}{\bf f} also shows that local disorder in cubic CsPbCl$_3$ leads to large electron and hole effective mass 
enhancements compared to the monomorphous structure, as well as to a linear, smooth increase of the effective masses 
with temperature across phase transitions. 
Such an enhancement in effective mass is expected to significantly impact the calculation of carrier mobilities.

\begin{figure}[t!]
\includegraphics[width=0.55\textwidth]{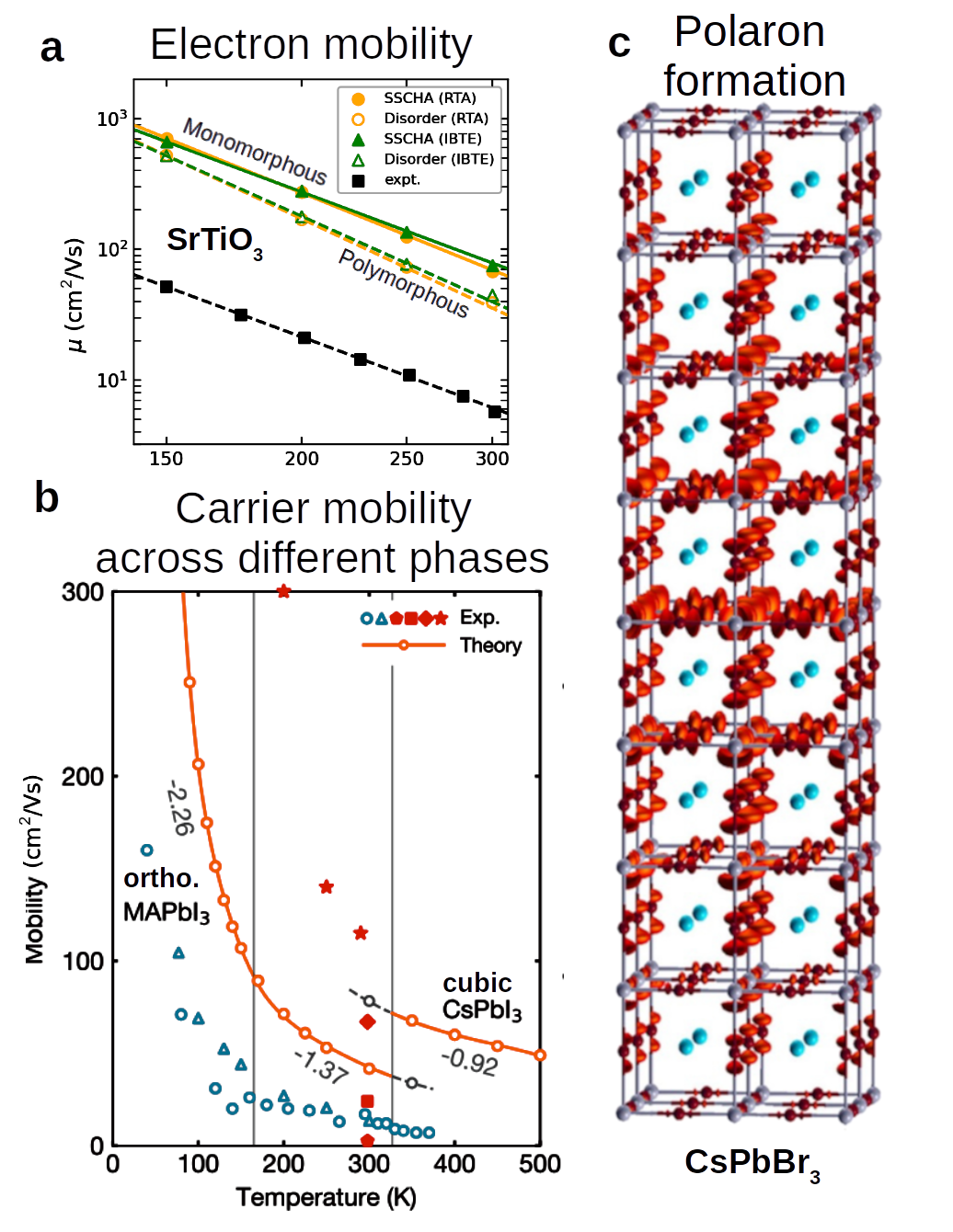}
\caption{\label{fig4} {\bf Impact of local disorder on transport and polaron formation.} 
(a) Calculated electron mobility of monomorphous and polymorphous cubic SrTiO$_3$ compared with experiment. 
Reproduced with permission from Ref.~\cite{Ranalli2024}, copyright (2024) by the American Physical Society.
(b) Calculated average charge-carrier mobility of orthorhombic MAPbI$_3$ and cubic CsPbI$_3$ compated with experiment. 
Reproduced with permission from Ref.~\cite{Ponce2019}, copyright (2019) by the American Chemical Society.
(c) Polaron charge density in CsPbBr$_3$ calculated by PBE0 in a supercell after adding a negative 
charge. Figure adapted from Ref.~\cite{Miyata2017}, copyright (2017) by the American Association for the Advancement 
of Science.}
\end{figure}

Figure~\ref{fig4}{\bf a} compares calculated electron mobilities, obtained using monomorphous and polymorphous structures of 
SrTiO$_3$, with experimental data as reported in Ref.~\cite{Ranalli2024}. The mobilities are computed using perturbative methods 
based on the relaxation time approximation (RTA) and iterative solutions of the Boltzmann transport equation (IBTE), as 
implemented in the {\tt EPW}~\cite{Lee2023} code. Anharmonic phonons are obtained via SSCHA. 
The effect of local disorder is incorporated by substituting the electronic eigenvalues computed 
for the monomorphous structure with those obtained from the polymorphous structure, while retaining the wavefunctions 
of the monomorphous configuration. 
Despite this approximation, local disorder reduces the electron mobility by 43\% at 300K, significantly 
improving agreement with experiment. An even greater reduction is achieved when local disorder is combined 
with electron-phonon coupling effects on the electronic structure, lowering the initial mobility overestimation from 
838\% to 290\%~\cite{Ranalli2024}, a remarkable improvement.
Zhou and Bernardi~\cite{Zhou2019} using the {\tt Perturbo} code~\cite{Zhou2021} achieved excellent agreement with experimental 
electron mobilities of SrTiO$_3$ by combining the retarded cumulant approach with the Kubo formula, capturing both coherent 
quasiparticles and incoherent polaronic effects through the spectral function. 
While their method treats dynamical electron-phonon interactions accurately, it neglects local structural disorder. 
Combining these two effects remains an open and promising direction for future research, particularly in understanding their 
joint impact on charge transport~\cite{Quan2024}. 
Ponc\'e et al~\cite{Ponce2019} investigated the origin of low carrier mobilities in halide perovskites 
by performing {\tt EPW} calculations within the RTA for the orthorhombic phase of MAPbI$_3$ and cubic phase of CsPbBr$_3$. 
Their results, presented in Figure~\ref{fig4}{\bf b}, illustrate the influence of local disorder on carrier mobilities 
in an indirect yet insightful manner. Employing the orthorhombic structure, which incorporates the effect of octahedral
tilting similar to local disorder in the cubic phase, yields better agreement with 
experimental data~\cite{Milot2015,Karakus2015}. This holds even for temperatures above 300~K, 
where the tetragonal to cubic phase transition takes place. Instead, using monomorphous cubic CsPbI$_3$ leads to 
a significant increase of the calculated carrier mobilities, leading to the overestimation of the experimental data 
by a factror of 7.

Figure~\ref{fig4}{\bf c} shows the calculated charge density of a large hole polaron in cubic CsPbBr$_3$ 
using a 2$\times$2$\times$8 supercell. Polaron is a manifestation of electron-phonon coupling, 
where the carrier distorts the surrounding atomic structure and becomes dressed by phonons, forming a 
quasiparticle. The calculations were performed by removing one electron and 
performing a geometry relaxation using the hybrid PBE0 functional to explain large polarons 
observed experimentally using time-resolved optical Kerr effect spectroscopy~\cite{Miyata2017}.
However, the computational approach employs a monomorphous cubic structure as the 
reference when introducing a charge and analyzing structural distortions. 
When a charged system is compared only to an idealized reference, the resulting lattice distortions may be overestimated 
or misattributed entirely to polaron formation. 
In reality, much of the distortion may already exist in the neutral, uncharged structure if modeled 
with a more physically accurate, symmetry-broken polymorphous configuration. 
Failing to use a locally disordered structure can lead to misleading conclusions about the nature, size, and stabilization 
energy of polarons in soft, anharmonic materials like CsPbBr$_3$.

\section{Future directions}

The findings discussed in this Perspective establish the importance of incorporating local disorder through polymorphous 
structural models to accurately predict and understand the electronic, vibrational, and electron-phonon properties of soft, 
anharmonic high temperature phases of materials. Despite the progress, the theoretical treatment of electron-phonon coupling 
in the presence of positional polymorphism remains at an early stage, offering numerous opportunities for future research.

One promising avenue is the integration of polymorphous supercells with state-of-the-art methodologies for electron-phonon coupling, 
such as the use of perturbative approaches implemented in state-of-the-art codes like {\tt EPW}~\cite{Lee2023} 
and {\tt Perturbo}~\cite{Zhou2021}. These methods typically assume high symmetry reference structures, limiting 
their predictive power in materials where symmetry breaking is intrinsic. 
Developing workflows  
with locally disordered configurations could significantly improve accuracy 
in modeling carrier lifetimes, charge transport, and phonon-assisted optical absorption.  
Regarding charge transport, another promising direction is to advance non-perturbative supercell approaches 
for anharmonic electron-phonon coupling, allowing the evaluation of carrier mobilities that include 
temperature-dependent band gap renormalization on top of local disorder effects~\cite{Quan2024}. 
An advantage of supercell approaches, despite lacking the elegance of fine Brillouin zone sampling, 
is their ability to seamlessly treat anharmonicity and electron-phonon coupling on the same footing, 
without relying on PES derivatives along harmonic phonons~\cite{Zhou2019}.

Another exciting direction lies in exploring non-equilibrium carrier dynamics in disordered lattices. First-principles 
simulations of carrier-phonon and phonon-phonon interactions in the ultrafast regime~\cite{Caruso2021}, 
can be extended to polymorphous structures to probe transient phenomena such as hot carrier relaxation, carrier localization, 
and coherent phonon generation, as well as to uncover scattering channels that are normally forbidden by symmetry 
in the monomorphous case. In parallel, excitonic effects in materials with strong local disorder remain largely 
unexplored; investigating how positional polymorphism influences exciton binding energies, lifetimes, and 
dissociation dynamics could offer new insights into light-matter interactions and optical response.  
Similarly, Auger recombination and radiative carrier lifetimes, which are central to 
optoelectronic device performance~\cite{Shen2018}, can be re-evaluated. 

The application of these frameworks to a broader class of materials, including layered/2D perovskites,  
double perovskites, antiperovskites, and quantum dots is another frontier. These systems often feature pronounced anisotropy and molecular 
complexity, making them ideal candidates for a polymorphous description. Particularly, the interplay of molecular cation dynamics 
with inorganic octahedral tilting and quantum confinement, deserves detailed investigation, potentially requiring the development 
of hybrid methodologies capable of treating the different sources of local disorder consistently~\cite{Zacharias2025b}.

It will also be interesting to compile databases of known and predicted anharmonic materials and conducting 
high-throughput first-principles studies to systematically explore their polymorphous structural networks across 
a broad chemical space~\cite{Jakob2025_disorder}. This effort would lay the groundwork for identifying trends and guiding 
the discovery of materials where local disorder plays a functional role. Moreover, given the large configurational space associated 
with polymorphous networks, machine learning force fields offer a powerful tool to accelerate structure exploration in 
large supercells and thus capture long range disorder as well as perform classification and even optoelectronic 
property prediction.  This effort could also provide insight into how polar symmetry breaking is entangled with local disorder 
in complex material systems, including but not limited to Ge-based perovskites.


It is worth noting that the polymorphous framework has certain limitations. These include the lack of an explicit treatment of 
relaxation processes detectable by low-energy spectroscopies and the restricted supercell sizes, which may underestimate correlation 
lengths and produce simulated Bragg-like features instead of experimental diffuse scattering. Furthermore, in its 
current implementation, local disorder is explored at fixed lattice parameters, neglecting strain-structure coupling and 
potentially biasing the accessible minima. Future extensions should incorporate variable-cell 
relaxations and larger supercells. Another limitation of the polymorphous framework is its quasi-static, Born-Oppenheimer treatment, 
which neglects non-adiabatic forces which can govern photoinduced structural dynamics~\cite{delaPeaMuoz2023}.

The full integration of the SCP theory with polymorphous structures offers a rigorous 
path toward modeling strongly anharmonic vibrations and demonstrate further the breakdown of the phonon quasiparticle picture. 
Such a framework would enable temperature-dependent phonon dispersions, lifetimes, thermal conductivity, 
and electron-phonon coupling coefficients to be computed on disordered environments, resolving current ambiguities in 
interpreting temperature-induced property changes. Modeling thermal conductivity in locally disordered materials
is critical, since the mean free paths can be reduced down to interatomic distances~\cite{Bhui2025}, and scattering becomes 
dominated by diffuson-like modes. This opens a path toward predicting ultralow thermal conductivity in materials exhibiting 
 entropy-driven structural complexity and glass-like behavior, despite their crystalline nature~\cite{Niedziela2020,Gupta2022,Wang2025}. 
Thermal conductivity relies 
on higher-order interatomic force constants, particularly third- and fourth-order terms governing phonon-phonon scattering, 
making machine learning force fields an essential tool for accelerating their computation in complex, disordered systems. 
Furthermore, extending local disorder to consider temperature-driven phase transitions and disorder-enabled functionalities 
such as ferroelectricity, spin polymorphism in paramagnetic phases, ionic transport, or polaron formation represents a rich area 
for computational development. 

From the experimental perspective, advancing characterization methods that are sensitive to local disorder is equally essential. 
While PDF analysis remains a powerful tool to probe local structure, complementary techniques such as high-temperature Raman 
spectroscopy, nuclear magnetic resonance measurements, inelastic neutron scattering for phonon DOS, and diffuse 
scattering can offer critical insights into vibrational dynamics and structural fluctuations in disordered 
systems~\cite{LaniganAtkins2021,Chengjie2025,He2024,He2025}. 
These methods provide 
additional observables that can be directly compared with first-principles predictions, allowing for more rigorous validation 
of polymorphous models, as well as elucidate long-range correlated disorder. Establishing such cross-validation pipelines between 
theory and experiment is key to uncovering the main mechanisms underpinning the efficiency of functional materials.

\section{Acknowledgments}
We acknowledge computational resources from the EuroHPC Joint Undertaking and supercomputer LUMI [https://lumi-supercomputer.eu/], hosted by CSC (Finland) and the LUMI consortium through a EuroHPC Extreme Scale Access call.

\bibliography{references}{} 
\end{document}